# Robust Surface States indicated by Magnetotransport in SmB$_6$ Thin Films

Jie Yong[1,2,*], Yeping Jiang[1,2], Xiaohang Zhang[1,2], Jongmoon Shin[3], Ichiro Takeuchi[1,3], Richard L. Greene[1,2]

[1]Center for Nanophysics & Advanced Materials, University of Maryland, College Park, Maryland 20742, USA

[2]Department of physics, University of Maryland, College Park, Maryland 20742, USA

[3]Department of Materials Science & Engineering, University of Maryland, College Park, Maryland 20742, USA

SmB$_6$ has been predicted and verified as a prototype of topological Kondo insulators (TKIs). Here we report longitudinal magnetoresistance and Hall coefficient measurements on co-sputtered nanocrystalline SmB$_6$ films and try to find possible signatures of their topological properties. The magnetoresistance (MR) at 2 K is positive and linear (LPMR) at low field and becomes negative and quadratic at higher field. While the negative part is known from the reduction of the hybridization gap due to Zeeman splitting, the positive dependence is similar to what has been observed in other topological insulators (TI). We conclude that the LPMR is a characteristic feature of TI and is related to the linear dispersion near the Dirac cone. The Hall resistance shows a sign change around 50 K. It peaks and becomes nonlinear at around 10 K then decreases below 10 K. This indicates that carriers with opposite signs emerge below 50 K. Two films with different geometries (thickness and lateral dimension) show contrasting behavior below and above 50K, which proves the surface origin of the low temperature carriers in these films. The temperature dependence of magnetoresistance and the Hall data indicates that the surface states are likely non-trivial.

*Email: jyong@umd.edu

With spin-momentum locked surface states which are robust against impurity scatterings, topological insulators (TIs) are promising in potential spintronic and quantum computing applications [1-2]. Unfortunately the bulk of most well-known topological insulators, such as $Bi_2Se_3$ or $Bi_2Te_3$, are not insulating due to lattice vacancies or self-doping [3]. This makes it difficult to utilize the surface properties of TIs because a conducting bulk complicates the topological surface transport. The prediction [4-5] and discovery [6-12] of $SmB_6$ as a topological insulator is extremely important because $SmB_6$ has a truly insulating bulk state. Moreover, it is well known that $SmB_6$ is also a Kondo insulator which means there might be an interesting interplay between correlated physics and topological properties. The fundamental step of utilizing these topological and correlated properties in actual devices requires preparation of $SmB_6$ in thin film forms. In our previous work [11], we have shown how we can grow $SmB_6$ nanocrystalline films by co-sputtering $SmB_6$ and B targets and demonstrated their basic physical properties. It would be interesting to know how disorder, which is typical in these films, affects the believed robust topological surface states. For the present work, we have measured electrical transport of the $SmB_6$ films in magnetic field, in both longitudinal and transverse directions. Although neither weak-antilocalization nor any quantum oscillations are observed, the observation of linearly positive magnetoresistance (LPMR) and the peculiar Hall signal lead us to conclude that our data are consistent with the existence of a robust topological surface state at low temperatures.

A well-known Kondo insulator [13], upon decreasing the temperature, the resistivity of $SmB_6$ increases like an insulator but saturates at temperatures below 5 K. Recent transport measurements show that the resistance of this saturation is thickness-independent for both longitudinal [9] and transverse [6] directions. Moreover, weak antilocalization and linear magnetoresistance [7] have been observed in some single crystal $SmB_6$, and such results are used to support the presence of spin momentum locked surface states. It has been shown that doping $SmB_6$ with magnetic impurities diminishes this saturation, while non-magnetic impurities do not [12]. These facts suggest that the conduction is indeed from the surface state: it is protected by time-reversal symmetry, and is robust against non-magnetic scatterings. Quantum oscillations of the surface states have been observed by torque magnetometry measurements [8], and the tracking of the Landau levels in the infinite magnetic field limit points to -1/2, which indicates a 2D Dirac electronic state. Neutron [14-15] and surface sensitive measurements, such as angle resolved photoemission (ARPES) [16, 17] and scanning tunneling microscopy (STM) [18], confirm the formation of the hybridization gap and the existence of the surface states. One study [17], in particular, suggests that the surface states are spin polarized with spin momentum locking which is a signature of the topological states. Unfortunately, there are also reports that indicate the resistivity saturation is due to dangling surface bonds [19] or complicated by carbon impurities [20]. To test whether there are indeed topological surface states and how they behave under low mobility and high disorder, which are typical in our nanocrystalline films, we performed magnetotransport measurements, both in longitudinal and transverse directions and their temperature dependence in $SmB_6$ thin films. We show the signature of surface states in these magnetotransport measurements. From the temperature dependence and in comparison with studies on $SmB_6$ crystals, we argue that these surface states in our $SmB_6$ films are topologically non-trivial.

$SmB_6$ films are grown by co-sputtering of $SmB_6$ and B targets in ultrahigh vacuum at 800C. The films are annealed in situ at 800C for 3 hours, and wavelength dispersive spectroscopy (WDS) is used to make sure the samples we obtained are stoichiometric $SmB_6$. Details of fabrication and characterization are given in Ref. [11]. Both magnetoresistance and Hall coefficient measurements are carried out in a Quantum Design PPMS system with the field up to 9T and temperature down to 2K. The magnetoresistance and Hall data are symmetrized ($R'_{xx}(H)$ = [$R_{xx}(H) + R_{xx}(-H)$]/2) and antisymmetrized ($R'_{xy}(H)$ = [$R_{xy}(H) - R_{xy}(-H)$]/2), respectively, to avoid the mixing of the longitudinal and transverse signals.

Typical magnetoresistance at 300K, 50K and 2K, normalized to their zero field value, is shown in Fig. 1. At room temperature, there is not much MR observed. At 50K, the MR develops a negative parabolic dependence. This dependence has previously been reported and is attributed to the reduction of the hybridization gap from Zeeman splitting [21]. At 2K, an additional linearly positive MR is found for B < 4T. While the negative MR is seen by all the studies on $SmB_6$, the low-field MR varies significantly from sample to sample. For example, some studies do not observe this LPMR: instead they see a hysteric behavior, and it is interpreted as arising from a ferromagnetic order or a spin glass state on the surface [22-23]. Our overall MR is similar to one crystal study [7], although our overall MR change is two orders of magnitude smaller. This is because our films are nanocrystalline: the resistance is dominated by field-independent grain-boundary scatterings. This LPMR does not show any field angular dependence as seen in single crystals because the nano-grains have ramdom orientations. The fact that our data is qualitatively similar to this single crystal data indicates that our results are from the nanocrystalline regions of the films.

Usually non-magnetic metals have small, quadratic positive magnetoresistance, and it quickly saturates at some field [24]. Linearly positive magnetoresistance (LPMR) is unusual and has been observed in several systems such as multilayer graphene [25], semimetal Bi [26], doped semiconductor InSb [27], and other topological insulators [28]. It is usually a much larger effect which makes it more interesting in possible applications. It also saturates at a much larger field or does not saturate at all. To the best of our knowledge, there is no consensus on the origin of LPMR . To date, two mechanisms have been proposed to explain this unusual effect. First is the classical one proposed by Littlewood and Parish [29] which states that inhomogeneities in small gap semiconductors or semimetals create tails in both the conduction and valence bands in small-gap semiconductors causing them to overlap. Distorted current paths can misalign with the driving voltage and mix in the off-diagonal components of the magnetoresistivity tensor. As a result, the magnetotransport is dominated by the magnitude of the fluctuations in the mobility near the band edge. This has been demonstrated in intentionally doped InSb [27] and AgSe [30]. Another mechanism is a quantum one proposed by Abrikosov [31-32] where the conditions are much more stringent: first, individual quantum levels associated with the electron orbits must be distinct: $\hbar\omega > k_BT$, where $\omega$ is the cyclotron frequency and $T$ is the temperature. Secondly, the system must be in the "extreme quantum limit', where $\hbar\omega$ can exceed the Fermi energy $E_F$. Under this condition, electrons can coalesce into the lowest quantum state of the transverse motion in $H$. A large magnetic field, a very small effective mass, and also a small carrier density are needed to satisfy these conditions. To date, this condition has only been shown in some specially engineered semiconductors. [27] We do not believe the LPMR we observed satisfies these conditions because it is observed at relatively high temperatures and in small fields. The LPMR we observed might be in the classical scenario since the Kondo band gap in $SmB_6$ is only ~20meV and our nanocrystalline films are heavily disordered. Recently it has been demonstrated that the magnitude of this LPMR is highly correlated to the average mobilities of the carriers in another topological insulator $Bi_2Te_3$ films. [33] The small magnitude of our LPMR is consistent with the low carrier mobilities of our highly disordered films. It is worthwhile to note that most materials with linear band dispersion (for example near a Dirac cone) have a LPMR. This includes topological insulators $Bi_2Se_3$ [28, 34] which has a large band gap ~300meV and new materials such as $LaAgBi_2$ [35] $SrMnBi_2$ [36], and $Cd_3As_2$ [37]. Therefore it seems that LPMR is an intrinsic and universal property of 2-D topological surface states. More theoretical work on LPMR is needed to confirm this empirical observation.

Hall resistances as function of external magnetic field at different temperatures are shown in Fig. 2. At 50K and higher, $R_{XY}$ is linear with a negative slope. Below 40K, $R_{XY}$ becomes positive and non-linear. This behavior peaks at around 10K. Below 10K, $R_{XY}$ becomes smaller and more linear as the temperature is lowered to 2K. To quantify the linearity of the $R_{XY}(H)$, we introduce a parameter $\alpha = R_{XY}(4T)/4R_{XY}(1T)$ and plot it as function of temperature in Fig. 3. Above 40K, $\alpha$ is almost 1.00 indicating the linear behavior. Then its value decreases to 0.80 at 10K before it increases back up to above 0.90 at 2K. Non-linearity in Hall resistance usually indicates presence of two different types of carriers with different mobilities. This fact, together with the sign change, indicates that the low temperature carriers have opposite signs and a different origin from high temperature carriers.

In a simple one carrier model, the slope of the Hall resistance is related to the carrier density by $R_{XY}(H)/H = S/(ned)$ (Eq. 2), where n is the carrier density (negative for electrons and positive for holes), e is the electron charge, d is the film thickness, and S is the lateral geometric factor. The Hall slopes (in small H limit) as a function of temperature for two films are displayed in Fig. 4. Upon cooling down, the Hall coefficients of both samples show a sign change near 40K. Both peak at around 10K and then decrease at lower temperatures. These behaviors are qualitatively similar to those of single crystals: both the sign change and the small peak are also observed in single crystal samples [6, 38]. They are related to the opening of the hybridization gap at 40K and two carrier competitions at intermediate temperatures. The interesting point is the sign of our Hall data is opposite to that of the sign observed in crystals. (We checked our configuration carefully to make sure that our measured sign was correct.) We believe the sign discrepancy may be due to the semiconducting-like properties of $SmB_6$ resulting from very flat bands of f electrons near the Fermi level and the fact that the chemical potential of our films may lie differently from that of crystals. The modest increase of $R_{xy}$ from 40K to 10K that we observe is in direct contrast with that of crystals, where $R_{xy}$ increases by orders of magnitude. This is similar to the smaller longitudinal resistance $R_{xx}$ increase that we found previously [11], which is ascribed to a much larger surface-to-bulk ratio for our films (by orders of magnitude). Also due to the larger surface-to-bulk ratio, the surface state dominates the competition against the bulk states faster than in crystals in cooling down. For our films, the Hall signal was found to peak at 10K, compared to the peaking of around 5K in single crystals [12, 38].

To further confirm that the low temperature carriers are from the surface, we take the two samples with different geometries (different S and d in Eq. 2), and compare their Hall signals in Fig. 4. Interestingly, while Sample B has a larger $R_{xy}$ than Sample A above 40K, and below 40K, Sample B has a smaller $R_{XY}$ compared to that of Sample A. This is counter-intuitive because the only parameter in Eq (2) that depends on the temperature is the carrier density. Usually the geometric difference only changes the scale of the Hall signal but not the temperature dependence of the signal. Since both samples are chemically similar, we do not expect this reversal of $R_{xy}$ magnitudes if the signal is always from the bulk. This discrepancy can be explained by considering that the "carrier thickness" d has temperature dependence. At high T, d is the whole film thickness because bulk is conducting. At low T, if only the surface conducts, d is the surface layer thickness, independent of the actual film thickness. In other words,, at higher temperatures, the difference is from the difference in S/d but at lower temperatures the difference is from the difference in A only. This proves that the low temperature carriers are from the surface in our nanocrystalline films, as also found in single crystals [6].

We note that the existence of the surface conducting layer does not mean the surface states are topological. The lack of both weak antilocalization and quantum oscillations from surface states, which are observed in most other TIs [39-41], is perplexing. However, we believe this can be resolved by carefully checking the conditions under which these quantum phenomena occur. It has been shown that even the best single crystals of $SmB_6$ do not show any quantum oscillations from transport measurements up to 100T [21]. Since the bulk is a Kondo insulator, no oscillations should be expected from the bulk. For the surface states, a high carrier mobility is required to observe oscillations: this is given by a simple condition that μB >1 where μ is the mobility of the carriers. One recent study shows that the surface carrier mobility in $SmB_6$ crystals is only 133 $cm^2$/V/s [9], which makes μB no greater than about 0.1 (the thin film value would be even smaller due to disorder). This is too small for observing quantum oscillations. Only one torque magnetometry study [8] has reported an observation of quantum oscillations, but in their case, the sensitivity is enhanced by a $H^2$ term in the torque formula. Three unique frequencies have been identified in their oscillations. The angular dependences of these three oscillations all point to the surface as their origin. Extrapolation to the infinite field limit gives an intercept of near -1/2, verifying the Berry phase of π, a signature of topological surface states. Therefore, it is the low mobility of surface carriers which prevents the observation of quantum oscillations.

As for weak antilocalization, there is an ongoing debate over the low field intrinsic magnetoresistance behavior at lowest temperatures. It has been claimed that observation of dips at low field and at mK temperatures in some samples is an indication of weak antilocalization (WAL) [7]. On the other hand, the observed dips have been reported to display measurement sweep rate dependence, and the observed apparent WAL has been attributed to surface magnetism [22]. Yet some have instead observed hysteric behaviors and have also attributed them to quantum transport due to surface magnetic domains [23]. Even if the observed dips are indeed due to WAL, it has been found only in a small fraction of single crystal samples and only at mK temperatures. We have also cooled our films down to 20mK, but neither WAL nor any hysteric behavior was observed (data not shown). It seems that WAL, if it exists at all, is not a robust phenomenon for $SmB_6$, as least compared to other topological insulators such as $Bi_2Se_3$, where WAL is observed by many groups at much higher temperatures [39-41]. One reason behind this difference may be that that surface states of $SmB_6$ only form below 50K and the hybridization gap is only 20meV. This is more than an order of magnitude smaller than that of $Bi_2Se_3$, which is 300meV. The energy scale here is much smaller making the observation of WAL much more difficult in $SmB_6$.

Despite all the arguments above, we cannot rule out the possibly that our observed surface states are trivial. In fact there are some reports that indicate that the surface states originate from dangling boron bonds [19] or from carbon contaminations [20]. One angle-resolved MR measurements [43] indicates there are *both* trivial and topological surface states. It is well-known that the electronic potential of a semiconductor surface is usually different from that of the bulk resulting in an internal electric field can. In such an instance, the band bending effect might lead to a formation of a two-dimensional electron gas (2DEG). This has been found to be the case for $Bi_2Se_3$ [40-41], where a 2DEG and topological states coexist on the surface. However, if our observed surface state is trivial as in a 2DEG, it is hard to explain our observed magnetotransport data. Chemically formed 2DEG or trivial surface states usually do not display strong temperature dependence. If we observe some properties from these trivial surface states, we would observe it at a much wider temperature range. Instead our observed LPMR and Hall sign change occur only below 50K. This indicates that the observed surface state is strongly associated with the opening of the hybridization gap and is likely to be topological. To definitively distinguish the trivial surface state from the topological surface state, further studies involving fabrication of high quality samples and surface-sensitive measurements are needed.

In conclusion, we have measured magnetoresistance and Hall coefficients on $SmB_6$ nanocrystalline films down to 2K. A linear positive MR is identified at low temperatures and is suggested to be associated with the linear dispersion near a Dirac point. The sign change of the Hall coefficient and its nonlinearity at lower temperatures indicates the presence of a robust surface conduction channel at low temperatures. The temperature dependence of these phenomena indicates the surface states are bound to the opening of the hybridization gap and they are most likely topological.

This work was supported by ONR N00014-13-1-0635 and NSF DMR 1410665. J. Y. also thanks the Maryland NanoCenter for technical support.

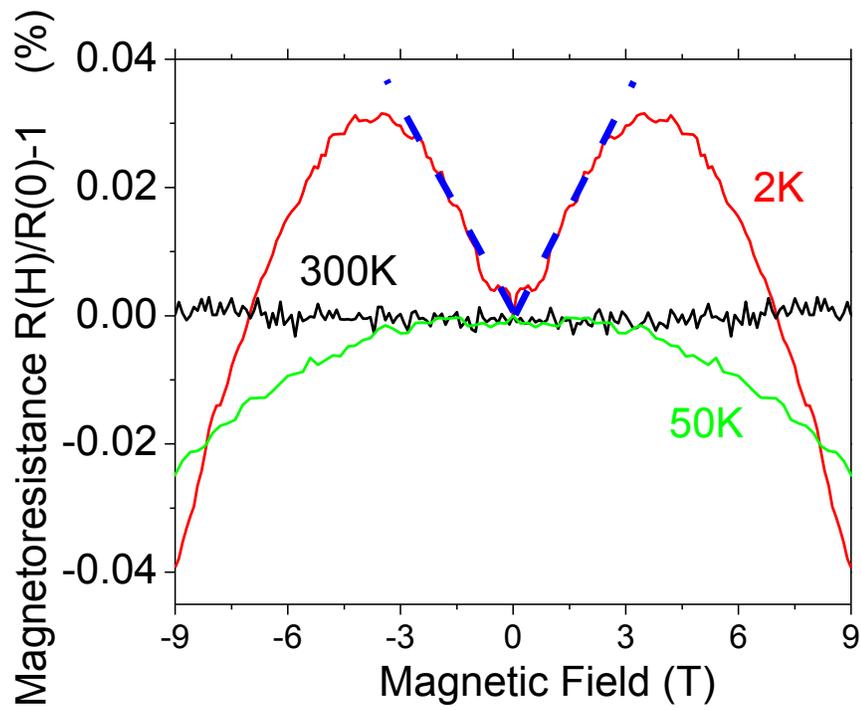

Figure 1: Magnetoresistance change in $SmB_6$ films defined by [R(H)/R(0)-1] *100% as a function of magnetic field at 300K (in black), 50K (in green) and 2K (in red). The blue dashed lines are linear guide to the eye.

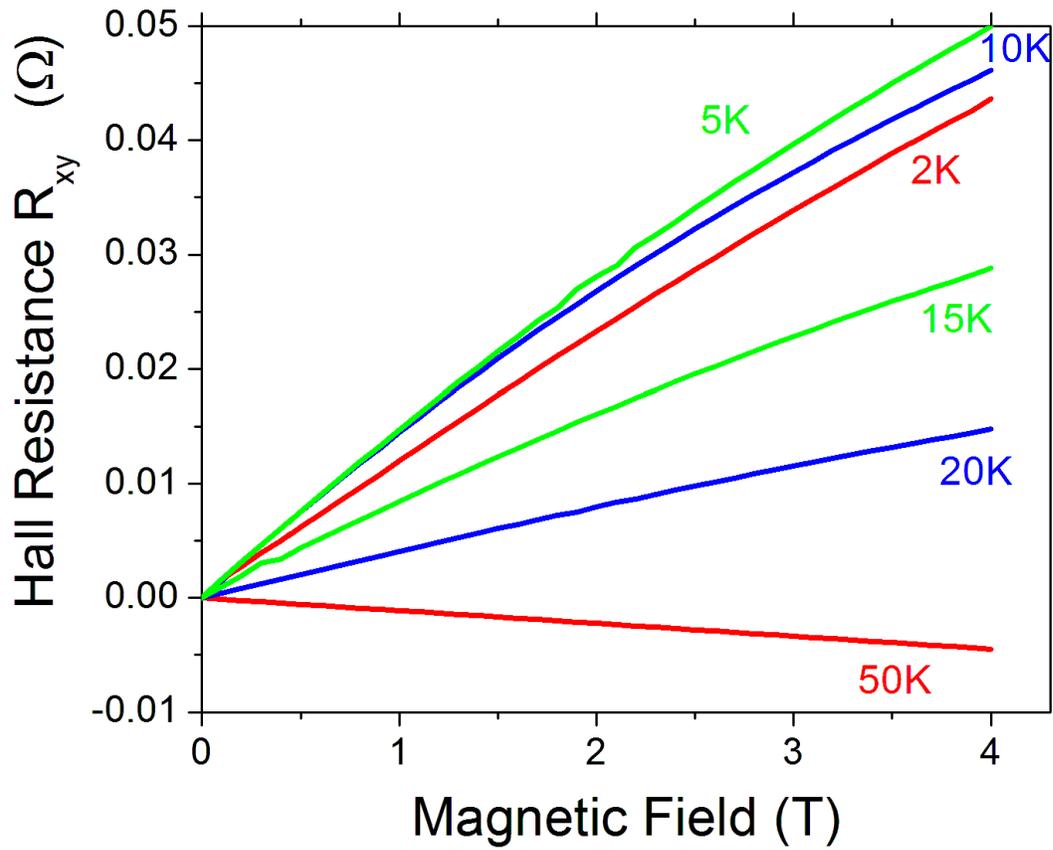

Figure 2: Hall resistance $R_{xy}$ of $SmB_6$ films as function of magnetic field at selected temperatures. The field is perpendicular to the film surface and the data is anti-symmetrized at zero field.

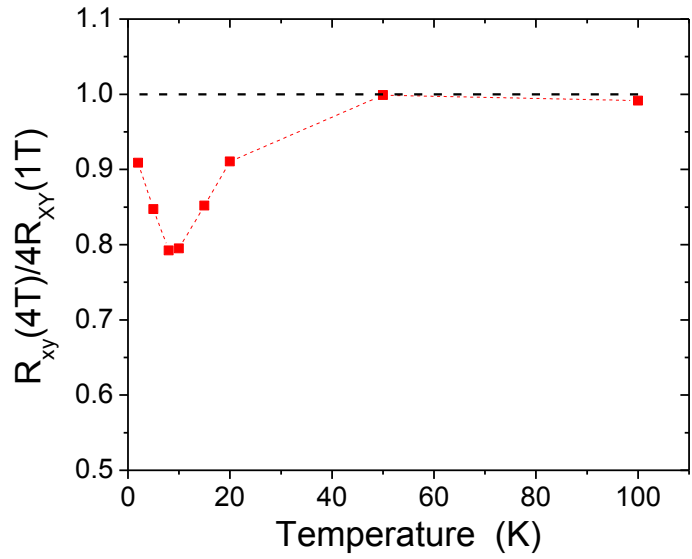

Figure 3: Hall resistance linearity parameter defined as $R_{XY}(4T)/[4*R_{XY}(1T)]$ as function of the temperature. The dashed black line shows the perfectly linear Hall resistance.

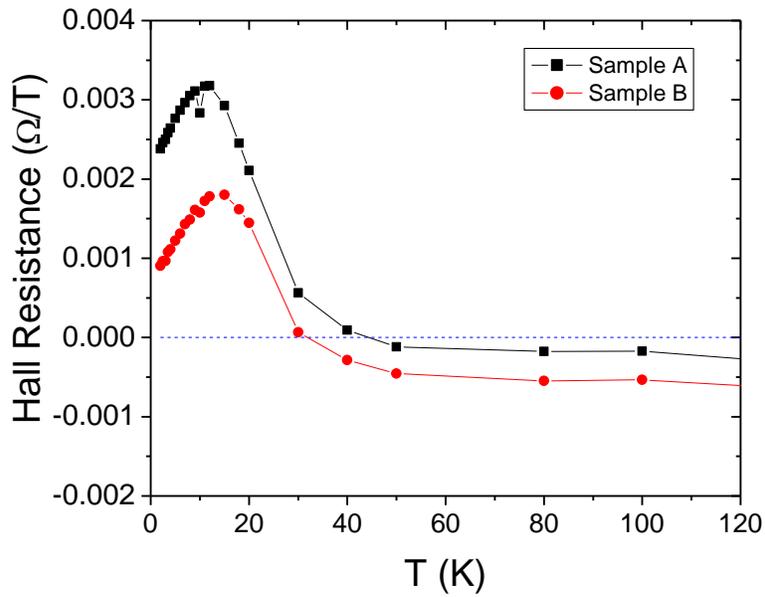

Figure 4: Hall resistance $R_{XY}/B$ as function of temperature for two samples different thickness and lateral geometries. The dashed line indicates where the Hall coefficient changes sign.

1. X. L. Qi, and S. C. Zhang, Rev. Mod. Phys. 83, 1057 (2011).

2. M. Hasan, and C. Kane, Rev. Mod. Phys. 82, 3045 (2010).

3 Y. Ando, J. Phys. Soc. Jpn. 82, 102001 (2013).

4. M. Dzero, K. Sun, V. Galitski, and P. Coleman, Phys. Rev. Lett. 104, 106408 (2010).

5. F. Lu, J. Zhao, H. Weng, Z. Fang, and X. Dai, Phys. Rev. Lett. 110, 096401 (2013).

6. D. J. Kim, S. Thomas, T. Grant, J. Botimer, Z. Fisk and Jing Xia, Surface Hall Effect and Nonlocal Transport in $SmB_6$: Evidence for Surface Conduction, Scientific Reports, 3, 3150 (2013).

7. S. Thomas, D.J. Kim, S. B. Chung, T. Grant, Z. Fisk and Jing Xia, Weak Antilocalization and Linear Magnetoresistance in The Surface State of $SmB_6$, arXiv:1307.4133 (2013)

8. G. Li, Z. Xiang, F. Yu, T. Asaba, B. Lawson, P. Cai, C. Tinsman, A. Berkley, S. Wolgast, Y. S. Eo, Dae-Jeong Kim, C. Kurdak, J. W. Allen, K. Sun, X. H. Chen, Y. Y. Wang, Z. Fisk and Lu Li, Two-dimensional Fermi surfaces in Kondo insulator $SmB_6$, Science, 346, 1208 (2014).

9. P. Syers, D. Kim, M. S. Fuhrer, and J. Paglione, Tuning bulk and surface conduction in topological Kondo insulator $SmB_6$, arXiv:1408.3402 (2014).

10. X. Zhang, N. P. Butch, P. Syers, S. Ziemak, R. L. Greene, and J. Paglione, Phys. Rev. X. 3, 011011 (2013).

11. Jie Yong, Yeping Jiang, Demet Usanmaz, Stefano Curtarolo, Xiaohang Zhang, Linze Li, Xiaoqing Pan, Jongmoon Shin, Ichiro Tachuchi and Richard L. Greene, Robust Topological Surface State in Kondo insulator $SmB_6$ Thin Films, Appl. Phys. Lett. 105, 222403 (2014).

12. D. J. Kim, J. Xia, and Z. Fisk, Nature Materials, 13, 466 (2014).

13. J. W. Allen, B. Batlogg, t and F. Wachter, Large low-temperature Hall effect and resistivity in mixed-valent SmB6, Phys. Rev. B 20, 4807 (1979).

14. W. T. Fuhrman, J. Leiner, P. Nikolic´, G. E. Granroth, M. B. Stone, M. D. Lumsden, L. DeBeer-Schmitt, P. A. Alekseev, J.-M. Mignot, S. M. Koohpayeh, P. Cottingham, W. Adam Phelan, L. Schoop, T. M. McQueen, and C. Broholm, Spin-exciton and topology in $SmB_6$, arXiv:1407.2647.

15. W. T. Fuhrman and P. Nikolic, In-gap collective mode spectrum of the Topological Kondo Insulator $SmB_6$, arXiv:1409.3220 (2014).

16. J. Jiang, S. Li, T. Zhang, Z. Sun, F. Chen, Z. R. Ye, M. Xu, Q. Q. Ge, S. Y. Tan, X. H. Niu, M. Xia, B. P. Xie, Y. F. Li, X. H. Chen, H. H. Wen and D. L. Feng, Nat. Commun. 4, 3010 (2013).

17. N. Xu, P. K. Biswas, J. H. Dil, R. S. Dhaka, G. Landolt, S. Muff, C. E. Matt, X. Shi, N. C. Plumb, M. Radovic, E. Pomjakushina, K. Conder, A. Amato, S. V. Borisenko, R. Yu, H.-M. Weng, Z. Fang, X. Dai, J. Mesot, H. Ding, and M. Shi, Nat. Commun. 5, 4566 (2014).

18. W. Ruan, C. Ye, M. Guo, F. Chen, X. Chen, G. Zhang, and Y. Wang, Phys. Rev. Lett., 112, 136401 (2014).

19. Z.-H. Zhu, A. Nicolaou, G. Levy, N. P. Butch, P. Syers, X. F. Wang, J. Paglione, G. A. Sawatzky, I. S. Elfimov, and A. Damascelli, Phys. Rev. Lett., 111, 216402 (2013).

20. W. A. Phelan, S. M. Koohpayeh, P. Cottingham, J.W. Freeland, J. C. Leiner, C. L. Broholm and T.M. McQueen, Correlation between Bulk Thermodynamic Measurements and the Low-Temperature-Resistance Plateau in $SmB_6$, Phys. Rev. X 4, 031012 (2014)

21. J. C. Cooley, C. H. Mielke, W. L. Hults, J. D. Goettee, M. M. Honold, R. M. Modler, A. Lacerda, D. G. Rickel, and J. L. Smith, High Field Gap Closure in the Kondo Insulator $SmB_6$, Journal of Superconductivity, 12, 171 (1999).

22. Yun Suk Eo, Steven Wolgast, Teoman • Ozt• urk, Gang Li, Ziji Xiang, Colin Tinsman, Tomoya Asaba, Fan Yu, Benjamin Lawson, James W. Allen, Kai Sun, Lu Li, Cagliyan Kurdak, Dae-Jeong Kim, Zachary Fisk, Hysteretic Magnetotransport in $SmB_6$ at Low Magnetic Fields, arXiv:1410.7430 (2014).

23. Y.Nakajima, P. Syers, X. F. Wang, R. X. Wang, and J. Paglione, One-dimensional edge state transport in a topological Kondo insulator, arXiv: 1312.6132 (2014).

24. Olsen, J. L. *Electron Transport in Metals* (Interscience, New York, 1962).

25. Adam L. Friedman, Joseph L. Tedesco, Paul M. Campbell, James C. Culbertson, Edward Aifer, F. Keith Perkins, Rachael L. Myers-Ward,‡ Jennifer K. Hite, Charles R. Eddy, Jr., Glenn G. Jernigan and D. Kurt Gaskill, Quantum Linear Magnetoresistance in Multilayer Epitaxial Graphene, *Nano Lett.*, *10*, 3962 (2010)

26. Yang, F. Y. *et al*. Large magnetoresistance of electrodeposited single-crystal bismuth thin films. *Science,* **284**, 1335–1337 (1999).

27. Jingshi Hu and T. F. Rosenbaum, Classical and quantum routes to linear magnetoresistance, *Nature Materials* 7, 697 (2008).

28. Xiaolin Wang, Yi Du, Shixue Dou, and Chao Zhang, Room Temperature Giant and Linear Magnetoresistance in Topological Insulator $Bi_2Te_3$ Nanosheets, Phys. Rev. Lett., 108, 266806 (2012)

29. Parish, M. M. & Littlewood, P. B. Non-saturating magnetoresistance in heavily disordered semiconductors. *Nature* **426**, 162–165 (2003).

30. Xu, R. *et al*. Large magnetoresistance in non-magnetic silver chalcogenides. *Nature* **390**, 57–60 (1997).

31. Abrikosov, A. A. Quantum magnetoresistance. *Phys. Rev. B* **58**, 2788–2794 (1998).

32. Abrikosov, A. A. Quantum linear magnetoresistance. *Europhys. Lett.* **49**, 789–793 (2000).

33. Z.H. Wang, L. Yang, X.J. Li, X.T. Zhao, H.L. Wang, Z.D. Zhang, Xuan and P. A. Gao, Granularity Controlled Non-Saturating Linear Magneto-resistance in Topological Insulator $Bi_2Te_3$ Films, *Nano Lett. 14* , 6510 (2014).

34. B. F. Gao, P. Gehring, M. Burghard and K. Kern, Gate-controlled linear magnetoresistance in thin $Bi_2Se_3$ sheets App. Phys. Lett., 100, 212402 (2012).

35. Kefeng Wang, D. Graf and C. Petrovic, Quasi-two-dimensional Dirac fermions and quantum magnetoresistance in $LaAgBi_2$, Phys. Rev. B 87, 235101 (2013)

36. Kefeng Wang, D. Graf, Hechang Lei, S. W. Tozer and C. Petrovic, Quantum transport of two-dimensional Dirac fermions in $SrMnBi_2$, Phys. Rev. B 84, 220401(R) (2011).

37. Tian Liang, Quinn Gibson, Mazhar N. Ali, Minhao Liu, R. J. Cava and N. P. Ong, Ultrahigh mobility and giant magnetoresistance in the Dirac semimetal $Cd_3As_2$, Nature Materials (2014)

38. J. W. Allen, B. Batlogg, and P. Wachter, Large low-temperature Hall effect and resistivity in mixed-valent $SmB_6$, Phys. Rev. B. 20, 4807 (1979).

39. J. Chen, H. J. Qin, F. Yang, J. Liu, T. Guan, F. M. Qu, G. H. Zhang, J. R. Shi, X. C. Xie, C. L. Yang, K. H. Wu,,Y. Q. Li, and L. Lu, Gate-Voltage Control of Chemical Potential and Weak Antilocalization in Bi2Se3, Phys. Rev. Lett., 105, 176602 (2010).

40. Namrata Bansal, Yong Seung Kim, Matthew Brahlek, Eliav Edrey and Seongshik Oh, Thickness-Independent Transport Channels in Topological Insulator $Bi_2Se_3$ Thin Films, Phys. Rev. Lett., 109, 116804 (2012)

41. Matthew Brahlek, Nikesh Koirala, Maryam Salehi, Namrata Bansal and Seongshik Oh, Emergence of Decoupled Surface Transport Channels in Bulk Insulating $Bi_2Se_3$ Thin Films, Phys. Rev. Lett., 113, 026801 (2014).

42. F. Chen, C. Shang, A. F. Wang, X. G. Luo, T. Wu, and X. H. Chen, Coexistence of nontrivial two-dimensional surface state and trivial surface layer in Kondo insulator $SmB_6$ arXiv:1309.2378 (2014).